\documentclass[12pt,preprint]{aastex}

\begin{document}

%
%--put definitions here
%
\def\msun{${\rm M_{\odot}} \;$}
\def\be{\begin{equation}}
\def\ee{\end{equation}}
\def\gcc{gcm$^{-3}$}
\def\bi{\begin{itemize}}
\def\ei{\end{itemize}}
\def\Mo{$M_{\odot}$}
\def\Lo{$L_{\odot}$}
\def\s{s$^{-1}$}

\title{Viscosity in cosmological simulations of clusters of galaxies}

\author{Marcus Br\"uggen\altaffilmark{1}, Mateusz Ruszkowski\altaffilmark{2,3}}

\altaffiltext{1}{International University Bremen, Campus Ring 1, 28759 Bremen, Germany}
\altaffiltext{2}{JILA, Campus Box 440, University of Colorado at Boulder, CO 80309-0440} 
\altaffiltext{3}{{\it Chandra} Fellow}

\begin{abstract}
The physics of the intracluster medium, in particular the values for
the thermal conductivity and the viscosity are largely unknown and
subject to an ongoing debate. Here, we study the effect of viscosity
on the thermal state of the intracluster medium using
three-dimensional cosmological simulations of structure formation. It
is shown that viscosity, provided it is not too far off from the
unmagnetised Spitzer value, has a significant effect on cluster
profiles. In particular, it aids in heating the cool cores of
clusters. The central cooling time of the most massive clusters in our
simulation is increased by more than an order of magnitude. In large
clusters, viscous heating may help to establish an entropy floor and
to prevent a cooling catastrophe.
\end{abstract}


\section{Introduction}

Cooling by bremsstrahlung and line emission leads to a loss of
pressure support in the centers of galaxy clusters, which, in the
absence of non-gravitational heating, would cause a slow, subsonic
inflow of gas towards the center of the gravitational well.  As a
result, cluster cores should cool and accrete gas at rates of hundreds
and more solar masses per year.  This scenario is in conflict with
observational evidence that indicates that mass deposition rates are
consistently lower than predicted. Moreover, the gas temperatures in
cluster centers are maintained typically above $\sim 2$ keV
\citep{peterson:01}.\\

More evidence for non-gravitational heating in clusters comes from
cluster scaling relations. These relations show departures from
self-similarity: In the absence of non-gravitational heating and
radiative cooling, the entropy is expected to scale linearly with the
mean cluster temperature. However, observations by Ponman, Sanderson
\& Finoguenov (2003), Pratt \& Arnaud (2005) and Piffaretti et
al. (2005) indicate a scaling of entropy roughly according to
$T^{2/3}$. Moreover, they reveal a systematic excess of entropy in
low-mass clusters (e.g., Ponman, Sanderson \& Finoguenov
2003). \cite{churazov:01,bruggen:02, bruggen:02a} have argued that
heating by a central AGN can keep the ICM from cooling dramatically in
the center.\\

Thermal conduction has been put forward to explain the absence of soft
X-rays from galaxy clusters (\cite{kim:03} and references
therein). Cosmological simulations with thermal conduction have been
performed by \cite{dolag:04} and \cite{jubelgas:04}. If thermal
conduction is at work, other transport processes such as viscosity are
bound to be important, too. Thermal conduction transports energy and
is mediated mainly by the faster electrons. Viscosity, on the other
hand, transports momentum and is mediated primarily by the more
massive ions. It is not clear that the suppression factors of both
transport processes should be the same. Whether they are the same may
depend on the scale magnetic fluctuations extend to. This scale may be
much larger than the gyroradii of electrons and ions (in which case
suppression factors could be comparable) or it could be comparable to
the ion gyroradius. The magnitude of the suppression factor is
motivated by various theoretical arguments (e.g., given in
\citealt{narayan:01}). However, we note that the precise value of the
suppression factor is highly uncertain and, depending on the nature of
magnetic turbulence, may even exceed the Spitzer value \citep{cho:03}
or be supressed well below it.\\

Based on observations of the Perseus cluster, it has been suggested by
\cite{fabian:03} that viscosity may play an important role in
dissipating energy injected by the central AGN. The case for this is
based on the existence of long, straight H$\alpha$-filaments that
appear to rule out the presence of strong turbulence in the cores of
galaxy clusters. The Reynolds number for a fluid flow whose viscosity
is suppressed with a factor, $f$, with respect to the Spitzer value,
is given by

\begin{equation}
%{\rm Re}\sim 50 \left(\frac{f}{0.3} \right )^{-1}\left(\frac{n_{\rm e}}{10^{-3}\ {\rm cm}^{-3}}\right ) \left(\frac{L}{1\ {\rm kpc}}\right ) \left(\frac{v}{500\ {\rm km\ s}^{-1}}\right )\left(\frac{T}{10^7\ {\rm K}}\right )^{-5/2} \ ,
{\rm Re}\sim 50\ f_{0.3}^{-1}\ n_{\rm e,-3}\ L_{\rm kpc}\ v_{500}\ T_7^{-5/2} \ ,
\label{eq:1}
\end{equation}
where $n_{\rm e,-3}$ is the electron number density in units of
$10^{-3}$ cm$^{-3}$, $L_{\rm kpc}$ the typical size of an eddy in
units of kpc, $v_{500}$ the associated velocity in units of 500 km
s$^{-1}$ and $T_7$ the temperature of the fluid in $10^7$ K. As such
values of Reynolds numbers are below the critical value that separates
laminar and turbulent regimes, it was concluded that viscosity can
play an appreciable role in the ICM, provided that viscosity is not
heavily suppressed.\\

Subsequently, heating by viscous dissipation of AGN-induced motions
has been simulated by
\cite{ruszkowski:04,ruszkowski:04b,bruggen:05}. It was concluded that,
provided viscosity is not suppressed significantly with respect to its
unmagnetised value, viscous dissipation of AGN-induced motions can
balance the radiative losses in the ICM. In \cite{reynolds:05} the
effect of viscosity on the evolution of radio bubbles was studied, and
it was found that viscosity had a stabilising effect on underdense
bubbles. \cite{fujita:04} have studied the dissipation of motions
induced by acoustic-gravity waves in cluster cores. They find that,
provided the wave amplitude is large enough, they can suppress the
radiative cooling of the cores. \cite{kim:05} have investigated the
heating of clusters by dynamical friction. They concluded that
friction can be an important supplier of heat but is unlikely to
prevent the onset of cooling flows.\\

Meanwhile, cosmological simulations of galaxy clusters with, both,
particle and grid-based methods have reached a fairly mature state. In
particular, the addition of increasingly sophisticated recipes for
radiative losses, star formation and stellar feedback have led to
cluster models that can produce many observed features of galaxy
clusters \citep{loken:02}.\\

\cite{motl:04} have simulated the formation of cool cores and observed
that any ``cooling flow'' is overwhelmed by the velocity field inside
the cluster, which has speeds of up to 2000 km s$^{-1}$. Nonetheless,
such violent motions did not prevent the formation of cool cores. This
shows that the formation of cool cores is inevitable unless some
source of heating is present. Full 3D cluster collision simulations by
\cite{ritchie:02} and \cite{ricker:01} show that mergers can disrupt
cooling flows. It is a common feature of all cosmological simulations
of galaxy clusters that the ICM shows a substantial velocity field
with many motions being supersonic. The ICM shows a complex dynamics
with cool fronts, filaments, shocks etc. New observations confirm this
picture. Detailed observations of unprecedented resolution by the
latest X-ray observatories have revealed a rich portfolio of
substructure in galaxy clusters. For example, \cite{schuecker:01} find
substructure in the majority of clusters in their REFLEX+BCS cluster
sample. The ubiquity of substructure points to a high frequency of
mergers and other events that prevent the cluster from relaxing to a
smooth, unperturbed state.\\

Both, observations and simulations suggest that the ICM is in violent
motion and that relaxed clusters, in which the gas sits almost
statically in its potential well, are very rare, if they exist at
all. In the presence of viscosity, a fraction of the kinetic energy in
these motions can be dissipated to heat the ICM. Thus, even in
inactive phases of a central AGN, there can be heating in the form of
viscous dissipation.

In this paper, we study the effect of viscosity on the intracluster
medium in a cosmological simulation. In particular, we wish to compute
to what extent viscous dissipation of random motions in a cluster
can contribute to the heating of cluster cores. In the next section,
we describe the technique and setup of our simulations. Finally, the
results are discussed in Sec.~\ref{sec:discussion}.

\section{Simulation}

The simulations were performed with the hydrodynamics code Enzo
developed by \cite{bryan:97} and \cite{norman:99}. Enzo is a
grid-based hybrid code that couples N-body computations with Eulerian
hydrodynamics. It is parallelised using the MPI (Message Passing
Interface) library, and is designed to perform simulations of
cosmological structure formation. In Enzo, the resolution of the
equations is adaptive in, both, time and space. In space, more finely
resolved child grids are produced where user-specified refinement
criteria are met. In our simulations, each child grid has a factor 2
higher spatial resolution than its parent grid and a factor 8 higher
mass resolution. For our purposes, the refinement is controlled by the
overdensity of dark matter and baryons, as well as the temperature. In
our setting, every cell whose overdensity was higher than 4 and had a
temperature above $10^6$ K was flagged for refinement. All cells on a
given level are advanced with the same time step.\\
 
Enzo uses a particle-mesh N-body method to calculate collisionless
particle dynamics. The dark matter particles are distributed onto the
grids using cloud-in-cell interpolation. The gravitational potential
is calculated on a periodic root grid using Fast Fourier
Transforms. To calculate the potential on the more finely resolved
subgrids, a multigrid relaxation method is used. Forces are computed
on the mesh by finite differences, and then interpolated onto particle
positions.\\

In our simulations, we included radiative cooling, assuming that all
species in the baryonic gas are in equilibrium. The cooling was
calculated from a tabulated cooling curve \citep{westbury:92} for a
plasma of fixed 0.3 solar abundance. The cooling curve is truncated at
a minimum temperature of $10^4$ K.

We used a star formation model by \cite{cen:92}. According to this
recipe, gas is converted into collisionless star particles in regions
that are Jeans unstable, contracting ($\nabla \cdot v < 0$) and
cooling rapidly. If these three conditions are met inside a
computational cell, a fraction of the cell's baryonic mass is
converted into a collisionless particle.\\

In \cite{oshea:05}, Enzo was compared to SPH simulations using the
GADGET code. General agreement in the distributions of temperature,
entropy and density within clusters was found between both numerical
methods.\\

Our simulations have a comoving box size of 64 Mpc/$h$. We chose a
standard flat $\Lambda$CDM cosmology with $\Omega_{\rm b}=0.044$,
$\Omega_{\rm m}=0.27$, $\Omega_{\Lambda}=0.73$, $h=0.71$ and
$\sigma_8=0.9$. We started from initial conditions at a redshift of
$z_{\rm init}=60$. With 8 levels of refinement and a root grid of
$64^3$ cells, we achieve an effective resolution of 3.9 kpc/$h$. The
simulations were run on 32 processor of an IBM p690 shared-memory
machine.\\

The evolution of internal energy is followed by solving the energy equation

\begin{equation}
%\rho\frac{de}{dt} = -p\Delta + \rho\dot{\epsilon}_{\rm visc} ,
\frac{\partial(\rho\epsilon)}{\partial t} + \nabla\cdot (\rho\epsilon+p)\mathbf{v} = \rho \mathbf{v}\cdot \mathbf{g} +\rho \dot{\epsilon}_{\rm visc} ,
\end{equation}
where $\epsilon$ is energy per unit mass, $\rho$ density, $\mathbf{v}$
velocity and $p$ pressure.  The dissipation of mechanical energy due
to viscosity, per unit mass of the fluid, is given by (Batchelor 1967,
Shu 1992, Landau \& Lifshitz 1997)

\begin{equation}
\dot{\epsilon}_{\rm visc}=\frac{2\mu}{\rho}\left(e_{ij}e_{ij}-\frac{1}{3}\Delta^{2}\right),
\end{equation}

\noindent
where $\Delta =e_{ii}$ and

\begin{equation}
e_{ij}=\frac{1}{2}\left(\frac{\partial v_{i}}{\partial x_{j}}+\frac{\partial v_{j}}{\partial x_{i}}
\right),
\end{equation}
and where $\mu$ is the dynamical coefficient of viscosity. In our
simulations with viscosity, we use a third of the standard Spitzer
viscosity for an unmagnetized plasma \citep{spitzer:62}, which is $\mu
= 6.0\times 10^{-17}T^{5/2}$ g cm$^{-1}$ s$^{-1}$, taking the Coulomb
logarithm to be $\ln \Lambda = 37$. For simplicity and want for any
better model, we assumed that $\mu$ is constant over cosmic time. This
is a simplification that is unlikely to be strictly true because
magnetic fields that inhibit viscosity will only be generated as
structures evolve in the universe. It is not clear when and how
magnetic fields form (see, e.g., \cite{bruggen:05}). However, at early
times, certainly at redshifts higher than 3, viscous heating is
relatively unimportant because the temperatures are quite
small. Consequently, it appears unlikely that this simplification will
affect our conclusions.\\

The timestep was modified to be the smaller value of the viscous and
the Courant time step, i.e.:

\begin{equation}
dt = {\rm min}(dt_{\rm Cour},dt_{\rm visc})\ ,
\end{equation}
where $dt_{\rm Cour}$ is the Courant time step and $dt_{\rm visc} =
0.1{\rm min}(dx_i^2/\mu)$.\\

Velocity diffusion was simulated by solving the momentum equation

\begin{equation}
\frac{\partial (\rho v_{i})}{\partial t}+
\frac{\partial}{\partial x_{k}}(\rho v_{k}v_{i})+
\frac{\partial P}{\partial x_{i}} = \rho g_{i}+
\frac{\partial\pi_{ik}}{\partial x_{k}} ,
\end{equation}
where

\begin{equation}
\pi_{ik}=\frac{\partial}{\partial x_{k}}\left[2\mu 
\left(e_{ik}-\frac{1}{3}\Delta\delta_{ik}\right)\right]
\end{equation}
and all other symbols have their usual meaning.

\section{Results and discussion}

\label{sec:discussion}

\begin{figure}[htp]
\begin{center}
\includegraphics[width=1.0\textwidth]{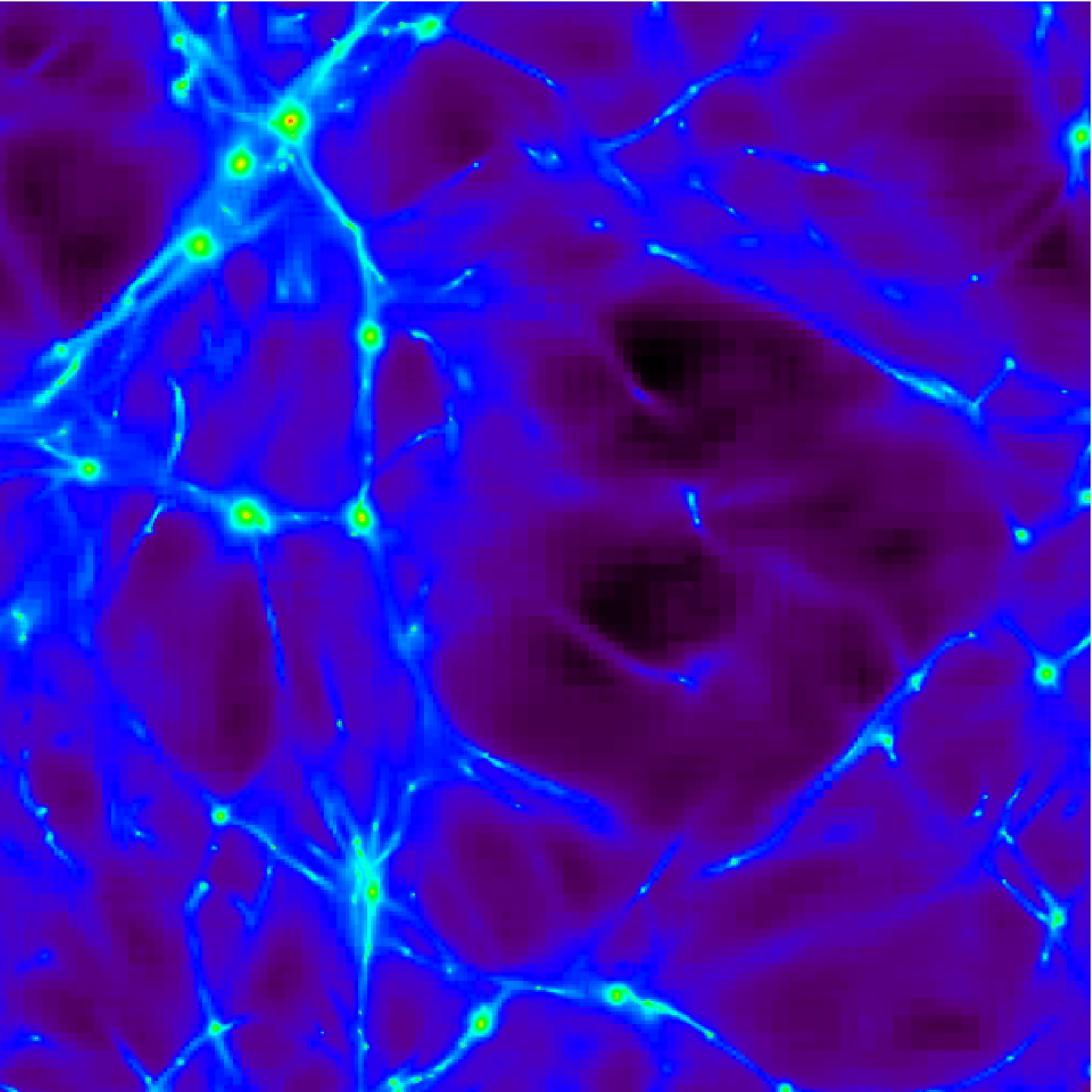}
\end{center}
\caption{Projected gas density at $z=0.25$ showing the most massive cluster in the simulation volume. The side length in this figure is 60/h Mpc.}
%from ENZOproj_dens_visc.jpg
\label{fig1}
\end{figure}

We produced two simulations: one without viscosity and one with a
third of Spitzer viscosity. Both runs included radiative cooling and
star formation. A density projection of the viscous simulation is
shown in Fig.~\ref{fig1}. The largest cluster is situated near the
upper edge of this figure. \\

Halos were identified using the HOP-algorithm developed by
\cite{eisenstein:97}. The parameters of the most massive cluster in
our simulation are summarised in table 1. The virial radius is
calculated for an overdensity of $\delta \rho/\rho =200$, and the
virial mass is the total mass (dark matter + baryons) within the
virial radius. $L_X$ is the total X-ray luminosity within $R_{\rm
vir}$ in the band from 0.1 - 2.4 keV. Finally, $\mathrm{M_{stars}}$ is
the total mass converted to stars. It is striking that, while the
gross features of the cluster, such as virial radius, mass and
temperature, are nearly identical between the runs, the X-ray
luminosities differ substantially. The X-ray luminosity depends on the
density squared and is thus dominated by the central portion of the
clusters. As discussed in the next section, there are pronounced
differences in the central densities, which do not affect the total
mass, though. Also, the mass converted to stars differ between the
two runs, with the viscous run producing less stars.\\

\begin{deluxetable}{| l || c | c | c}
      \tablewidth{0pt}
      \tablenum{1}
      \tablecolumns{3}
      \tablecaption{Properties of the most massive cluster in our sample\label{tab:clusters}}
      \startdata
      \hline
                            & without visc &  with visc      \\
      \hline
      \hline
      $\mathrm{R_{virial}}$
      & $1.11 \: \mathrm{Mpc}$
      & $1.12 \: \mathrm{Mpc}$ \\
%      & $0.73 \: \mathrm{Mpc}$ \\
      $\mathrm{M_{virial}}$
      & $2.2 \times 10^{14} \: \mathrm{M_{\odot}}$
      & $2.3 \times 10^{14} \: \mathrm{M_{\odot}}$ \\
%      & $5.82 \times 10^{13} \: \mathrm{M_{\odot}}$ \\
      $\mathrm{L_{x}}$
      & $1.6 \times 10^{46} \: \mathrm{erg} \, \mathrm{s^{-1}}$
      & $2.2 \times 10^{45} \: \mathrm{erg} \, \mathrm{s^{-1}}$ \\
%      & $9.7 \times 10^{43} \: \mathrm{erg} \, \mathrm{s^{-1}}$ \\
      $\mathrm{T_{\rm vir}}$
      & $3.0 \times 10^{7} \: \mathrm{K} $
      & $3.1 \times 10^{7} \: \mathrm{K} $ \\
%      & $1.2 \times 10^{7} \: \mathrm{K} $ \\
      $\mathrm{M_{stars}}$
      & $7.4\times 10^{12} \: \mathrm{M_{\odot}}$
      & $6.5\times 10^{12} \: \mathrm{M_{\odot}}$
%\hline
%      & 0.117 \\
      \enddata
\end{deluxetable}

%\begin{deluxetable}{| l || c | c | c}
%      \tablewidth{0pt}
%      \tablenum{1}
%      \tablecolumns{3}
%      \tablecaption{Properties of the two most massive clusters with viscosity\label{tab:clusters2}}
%      \startdata
%      \hline
%                            & 1 &  2   \\
%      \hline
%      \hline
%      $\mathrm{R_{virial}}$
%      & $1.37 \: \mathrm{Mpc}$
%      & $1.12 \: \mathrm{Mpc}$ \\
%%      & $1.04 \: \mathrm{Mpc}$ \\
%      $\mathrm{M_{virial}}$
%      & $3.78 \times 10^{14} \: \mathrm{M_{\odot}}$
%      & $2.05 \times 10^{14} \: \mathrm{M_{\odot}}$ \\
%%      & $1.95 \times 10^{13} \: \mathrm{M_{\odot}}$ \\
%      $\mathrm{L_{x}}$
%      & $7.3 \times 10^{44} \: \mathrm{erg} \, \mathrm{s^{-1}}$
%      & $1.3 \times 10^{45} \: \mathrm{erg} \, \mathrm{s^{-1}}$ \\
%%      & $4.3 \times 10^{44} \: \mathrm{erg} \, \mathrm{s^{-1}}$ \\
%      $\mathrm{T_{\rm vir}}$
%      & $4.2 \times 10^{7} \: \mathrm{K} $
%      & $2.8 \times 10^{7} \: \mathrm{K} $ \\
%%      & $2.48 \times 10^{7} \: \mathrm{K} $ \\
%      $\mathrm{f_{cool}}$
%      & 0.085
%      & 0.085 \\
%      & 0.117 \\
%      \enddata
%\end{deluxetable}

\begin{figure}[htp]
\begin{center}
\includegraphics[width=1.0\textwidth]{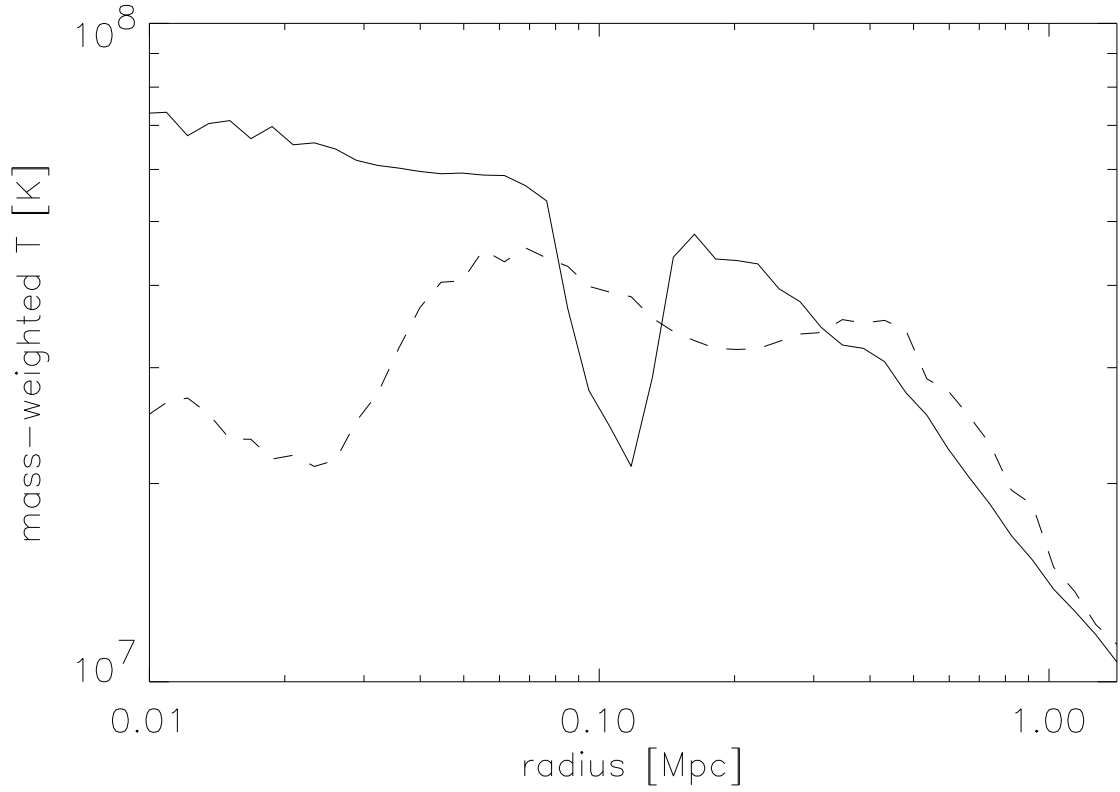}
\end{center}
\caption{Mass-weighted temperature profiles for the most massive cluster in our simulation (see table 1). The solid line corresponds to the run with viscosity, while the dotted line corresponds to the run without viscosity.}
\label{fig2}
\end{figure}

\begin{figure}[htp]
\begin{center}
\includegraphics[width=1.0\textwidth]{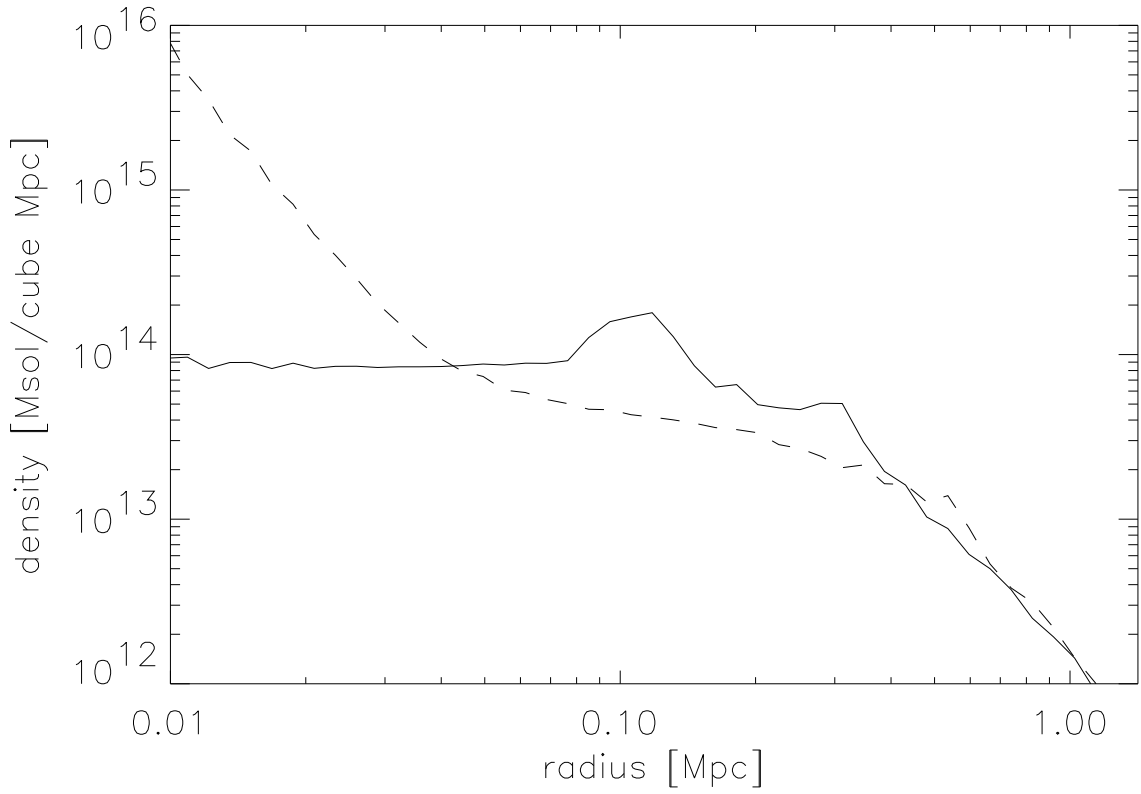}
\end{center}
\caption{Density profiles for the most massive cluster. The units are
$M_{\odot}$/Mpc$^{-3}$. The solid line corresponds to the run with
viscosity, while the dotted line corresponds to the run without
viscosity.}
\label{fig3}
\end{figure}

\begin{figure}[htp]
\begin{center}
\includegraphics[width=1.0\textwidth]{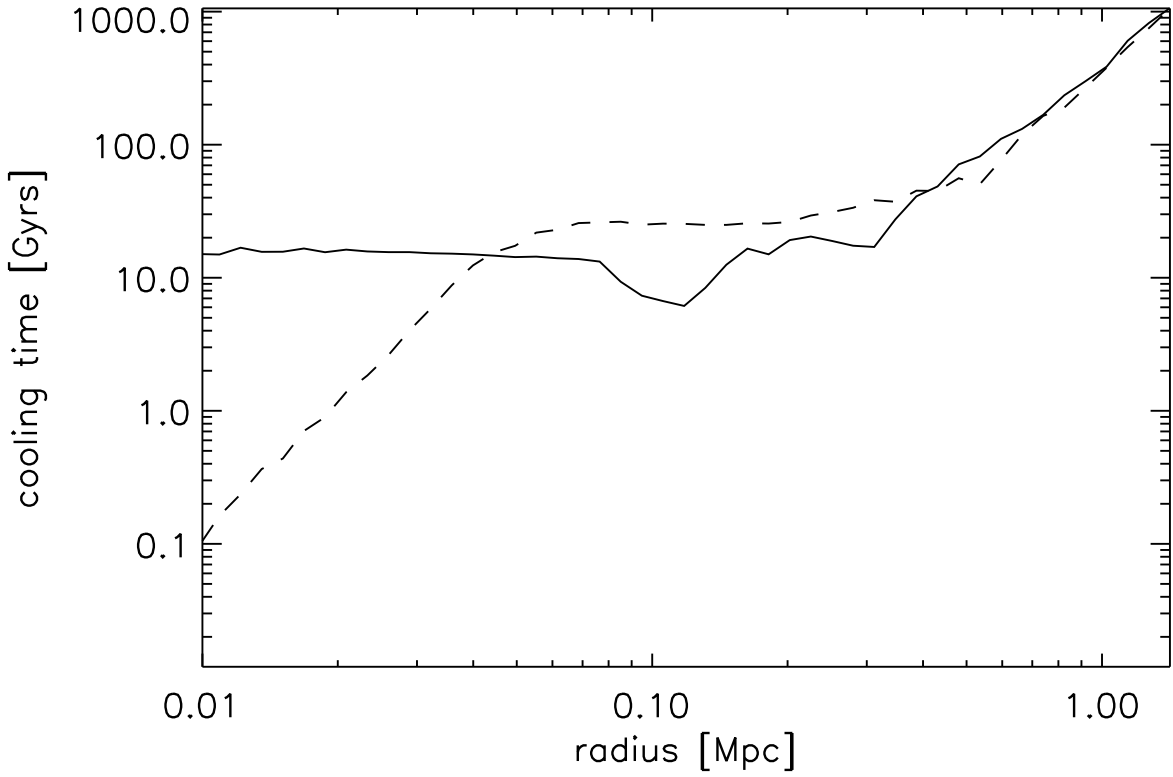}
\end{center}
\caption{Cooling time (in Gyrs) profiles for the most massive cluster. The solid line corresponds to the run with viscosity, while the dotted line corresponds to the run without viscosity.}
\label{fig4}
\end{figure}

\begin{figure}[htp]
\begin{center}
\includegraphics[width=1.0\textwidth]{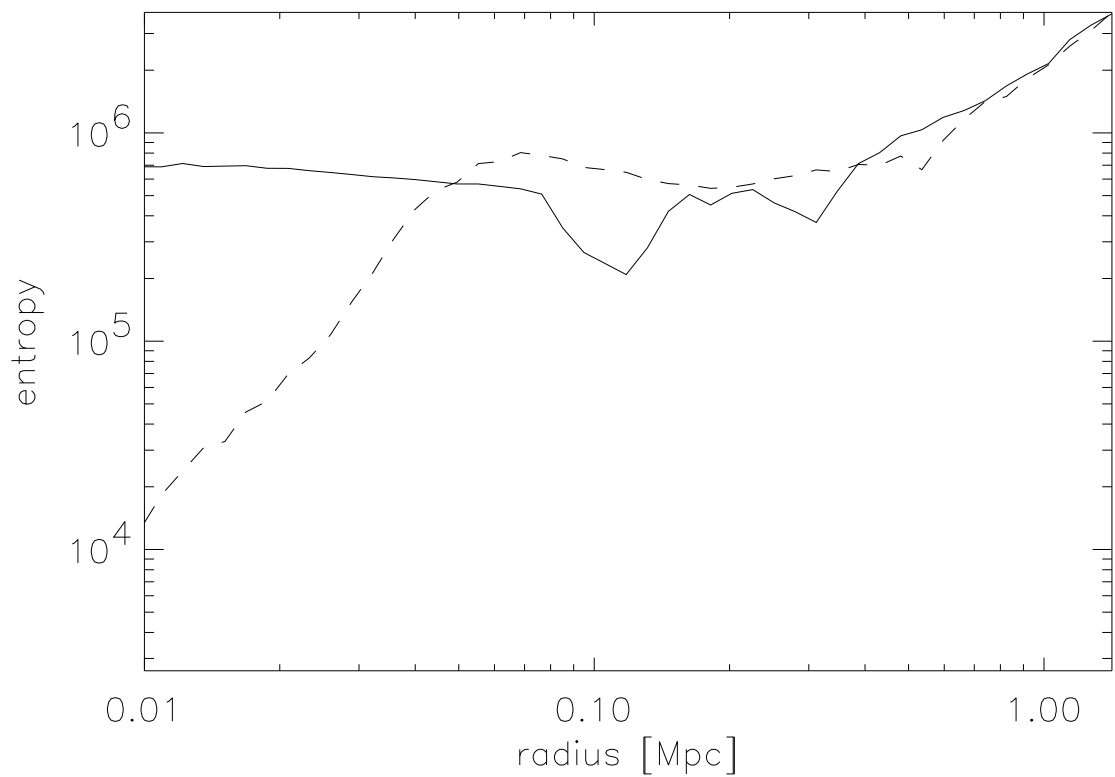}
\end{center}
\caption{Entropy profiles for the most massive cluster. The solid line
corresponds to the run with viscosity, while the dotted line
corresponds to the run without viscosity.}
\label{fig5}
\end{figure}

In Fig.~\ref{fig2} - \ref{fig5} we show the mass-weighted temperature,
density, cooling time and entropy, respectively, as a function of
radius in the most massive, non-merging cluster in our sample.  While
a cool core forms in the run without viscosity, it is absent in the
viscous run. As can be seen from the temperature profile,
Fig.~\ref{fig2}, viscosity has essentially removed the cool core and
the mass-weighted temperature even rises slightly in the centre. The
density profiles also show significant differences (see
Fig.~\ref{fig3}).  In the presence of viscosity, the density is nearly
constant in the core, whereas, in the non-viscous run, it rises sharply in
the inner 40 kpc.  With 1/3 Spitzer viscosity, the density is very
flat over the inner hundred kiloparsecs. In summary, we find that the
runs with viscous dissipation lead to a hotter and less dense
core. Consequently, the cooling time increases in the center, with
respect to the non-viscous runs (see Fig.~\ref{fig4}). The central
cooling time is about two orders of magnitude higher than in the
non-viscous case and larger than the Hubble time. If viscous heating
was this efficient, no other sources of heating would be required to
prevent a cooling catastrophe. The entropy, which is shown in
Fig.~\ref{fig5}, displays a very extended floor and has no central dip
as in the run without viscosity. A comparison of a statistically
relevant sample of simulated clusters can be compared with observed
samples, such as the one by \cite{donahue:05}. Thus, one may be able to
constrain the viscosity of the ICM from X-ray observations.\\

\begin{figure}[htp]
\begin{center}
\includegraphics[width=1.0\textwidth]{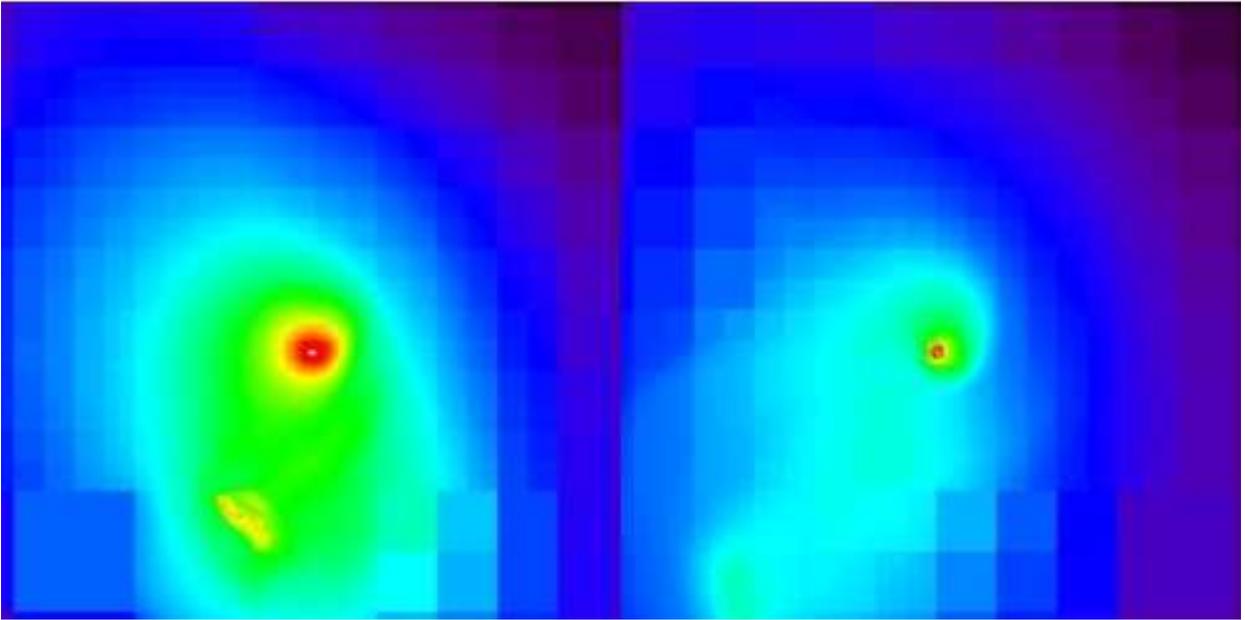}
\end{center}
\caption{X-ray flux of the most massive cluster normalised to the maximum. The left panel shows the cluster simulated with viscosity. The right panel shows the cluster simulated without viscosity.}
\label{fig6}
\end{figure}

In Fig.~\ref{fig6} we compare the X-ray luminosity from the cluster in
the band from 0.1 - 2.4 keV. In the case with viscosity, it is
apparent that the cluster is more extended and that it lacks a strong
emission spike in the center.\\

\begin{figure}[htp]
\begin{center}
\includegraphics[width=1.0\textwidth]{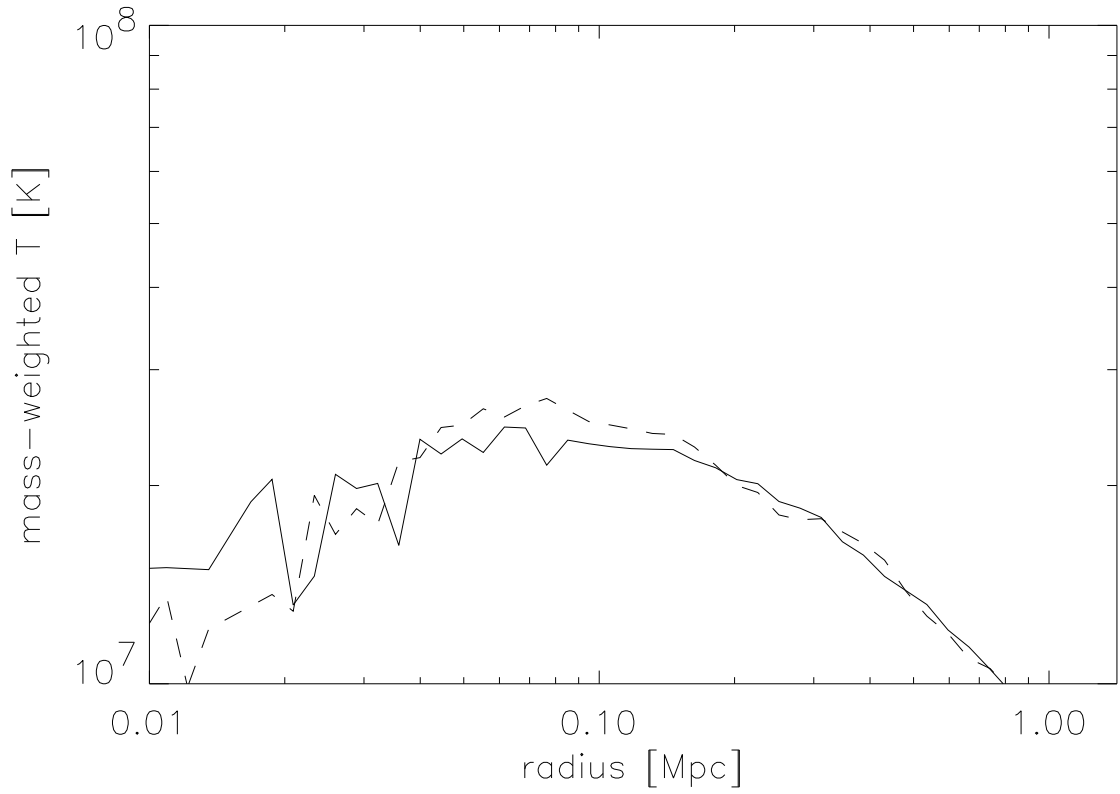}
\end{center}
\caption{Mass-weighted temperature profiles for a smaller cluster with a mass of $9.7\times 10^{13}\ M_{\odot}$. The solid line corresponds to the run with viscosity, while the dotted line corresponds to the run without viscosity.}
\label{fig7}
\end{figure}

\begin{figure}[htp]
\begin{center}
\includegraphics[width=1.0\textwidth]{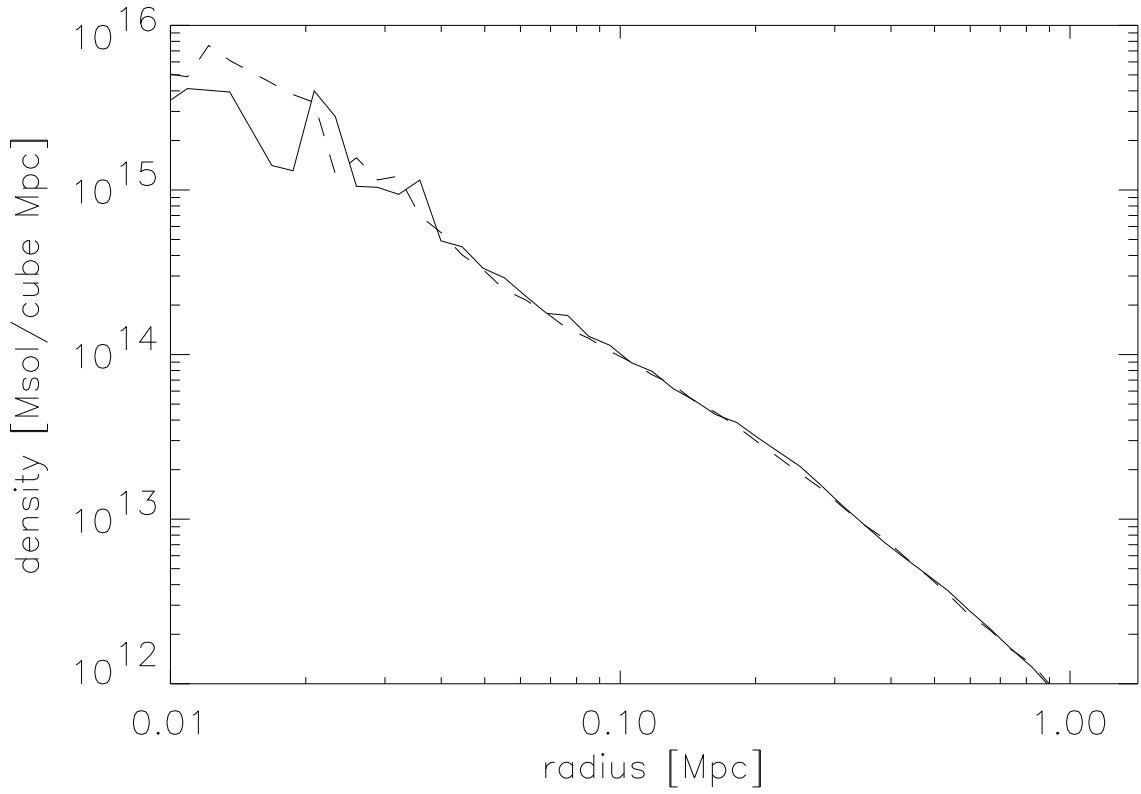}
\end{center}
\caption{Density profiles for a smaller cluster with a mass of
$9.7\times 10^{13}\ M_{\odot}$. The units are
$M_{\odot}$/Mpc$^{-3}$. The solid line corresponds to the run with
viscosity, while the dotted line corresponds to the run without
viscosity.}
\label{fig8}
\end{figure}

\begin{figure}[htp]
\begin{center}
\includegraphics[width=1.0\textwidth]{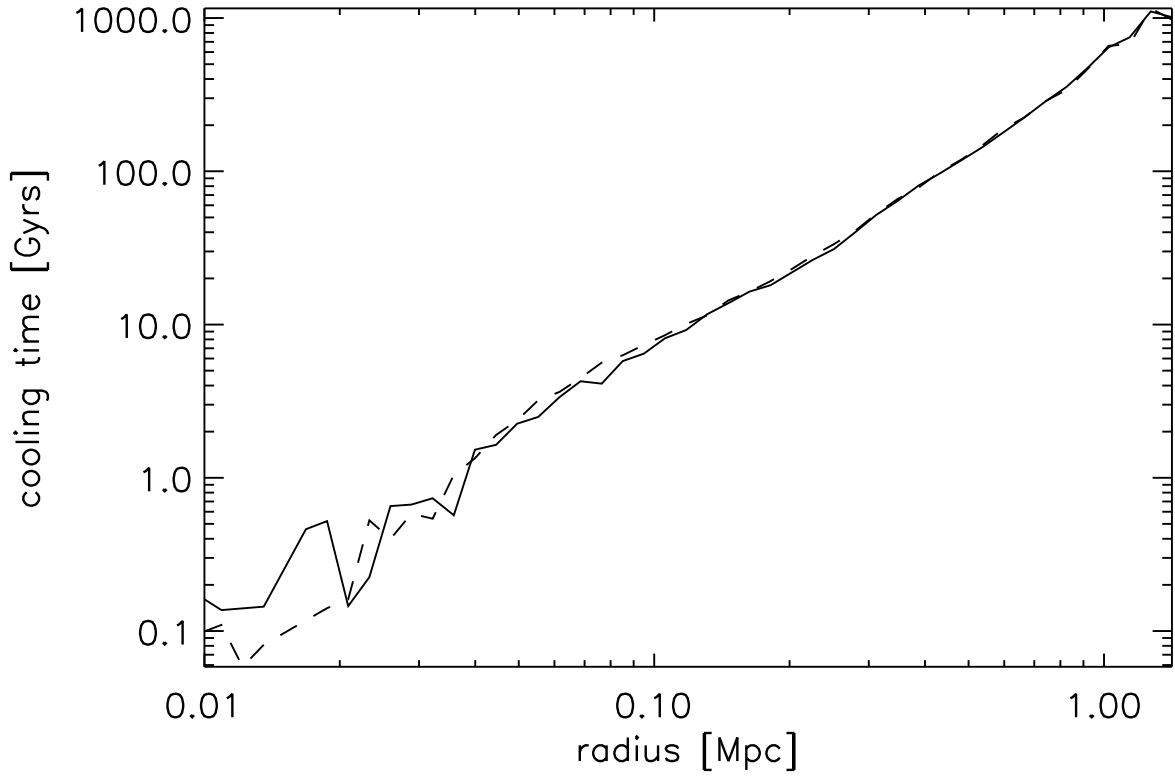}
\end{center}
\caption{Cooling time (in Gyrs) profiles for a smaller cluster with a mass of $9.7\times 10^{13}\ M_{\odot}$. The solid line corresponds to the run with viscosity, while the dotted line corresponds to the run without viscosity.}
\label{fig9}
\end{figure}

The temperature-dependence of viscosity biases this mode of heating
towards hotter, more massive clusters, and will, thus, affect cluster
scaling relations. We find that the effect of viscosity becomes
systematically less important with decreasing mass of the cluster. In
Fig.~\ref{fig7} - Fig.~\ref{fig9}, we show the corresponding profiles
of a smaller cluster with a mass of $9.7\times 10^{13}\
M_{\odot}$. Evidently, viscosity only has a minor effect on the
central density profile. A proper study of this effect requires a
bigger simulation box that contains a large number of clusters, and is
the subject of future work.\\

Note that in the simulations presented here, heat conduction has been
neglected. As argued above, heat conduction and viscosity are
intrinsically linked, and it would be interesting to study their joint
effect on the cluster. Since the suppression factors of viscosity and
conductivity are unknown, the consideration of heat conduction
introduces another, essentially free, parameter into the problem. For
this first simulation, we decided to study the isolated effect of
viscosity only. \\

In order to validate our code modules, we have repeated a low-level
test simulation in a small box with the FLASH code. The viscosity
routines in FLASH have been tested in
\cite{ruszkowski:04,ruszkowski:04b}. Starting from similar initial
conditions and including the same physics, we obtain very similar
results between Enzo and FLASH.

%In
%Fig.~\ref{fig7}, we show a slice through the computational box that
%displays the logarithm of the ratio between the viscous energy
%dissipated within one timestep over the total energy in the gas. The
%plot is centered on a filaments and shows clearly that most of the
%energy is dissipated in regions where the gas falls into halos.\\

As a result of numerical diffusion, the effective Reynolds number will
be finite, even in the zero-viscosity case. The effective Reynolds
numbers attainable in the simulation are proportional to the number of
grid points across the fluctuation of interest to the power $n$, where
$n=3$ is the order of the numerical scheme\footnote{See, e.g.,
\citet{bowers:91} for the definition of ``the order of the numerical
scheme'', as it is different from the customary definition of accuracy
of a perturbative calculation.}  \citep{porter:94}. As expressed in
Eq.~\ref{eq:1}, the Reynolds number in the ICM is of the order of 50,
assuming 0.3 of the Spitzer value. Test simulations with FLASH and
Enzo suggest that the effective Reynolds number for the resolution
chosen here is $>$ 1000. Both runs that we have presented here have
identical spatial resolution. Hence, the differences between the runs
are solely due to the physical viscosity.\\

Both, radiative losses and viscosity, are sensitive to the spatial
resolution of the computational grid. We experimented with different
refinement criteria and compared results from runs with different
effective resolutions. Generally, the effect of physical viscosity
becomes larger as the resolution increases. The same is true for
radiative losses. The results presented here seem to be reasonably
converged and differ only marginally from the run with a refinement
level less.\\

Assuming hydrostatic equilibrium when inferring cluster masses from
synthetic observations, leads to masses that are systematically lower
(by 10-15 per cent) than the actual cluster masses in
simulations. This affects cluster constraints on cosmological
parameters (see e.g., \citet{allen:04}). If the ICM is viscous, it is
conceivable that the assumption of hydrostatic equilibrium is better
met. Thus, gas viscosity may reduce systematic deviations of cluster
masses inferred from X-ray observations from their true values, which,
in turn, may have consequences for precision measurements of
cosmological parameters.

\section{Summary}

In a simple numerical experiment, we added physical viscosity to a
cosmological simulation and studied its effect on cluster
properties. The results of this experiment are interesting.\\

A viscosity not too far off from the unmagnetised Spitzer value can
alleviate the problem of cool cores in galaxy clusters. Viscosity
raises the temperature in the core and reduces the central cooling
times significantly. This effect is important for massive clusters,
but becomes ineffective for smaller clusters and groups. For 1/3 of
Spitzer viscosity, the effect is important for cluster above $\sim
10^{14}\ M_{\odot}$. Feedback from star and AGN has been ignored in
this work. These are additional sources of heat that will increase
viscous heating. The interplay of various feedback processes with a
possibly viscous ICM may complicate matters and should be looked at in
future work. \\

As discussed above, the actual value of the viscosity in the ICM is
subject to speculation, and the viscosity is essentially a free
parameter. Here, we assumed that the magnetic field is not too
efficient in suppressing viscosity and took a fiducial value of a
third of Spitzer viscosity. Value cited in various studies in the
recent literature assume value between 1/3 and 1/10 of the Spitzer
value. Only if magnetic fields should suppress the viscosity by 2
orders of magnitude or more, the effect will start to be
negligible. We defer the study of the effects of viscosity on the
statistical properties of clusters and groups to a future publication
based on a bigger simulation box containing a large number of
clusters.\\

Real clusters may possess complex magnetic field configurations that
lead to a very inhomogeneous and anisotropic viscosity. This, in turn,
can lead to complex dynamics of the ICM. It will be a challenge for
future work to disentangle the effects of transport processes and
magnetic fields.

%\noindent {\sc acknowledgement }

\acknowledgements{

We thank Brian O'Shea for providing us with non-public modules of Enzo
and for valuable help. Eric Hallman is thanked for insightful
discussions. M.B. acknowledges support by DFG grant BR 2026/3 within
the Priority Programme `Witnesses of Cosmic History' and the
supercomputing grant NIC 1658 at the John-Neumann Institut at the
Forschungszentrum J\"ulich. M.R. acknowledges support from NSF grant
AST 03-07502 and NASA through Chandra Fellowship award PF3-40029
issued by the {\it Chandra X-Ray Observatory} Center, which is
operated by the Smithsonian Astrophysical Observatory for and on
behalf of NASA under contract NAS8-39073.}

\newpage

%\bibliography{radio}
\bibliographystyle{apj}

%\clearpage

\end{document}